\documentclass[twocolumn,twoside,slac_two]{revtex4}
\usepackage{graphicx}
\usepackage{fancyhdr}
\newcommand{\roughly}[1]{\mathrel{\raise.3ex\hbox{$#1$\kern-0.85em
\lower1ex\hbox{$\sim$}}}}

\newcommand{\lsim}{\roughly<}

\def\beq{\begin{equation}}
\def\eeq{\end{equation}}

\begin{document}

\title{MSSM flat direction inflation}

\author{Rouzbeh Allahverdi$^{1,2}$}
\affiliation{$^{1}$ Perimeter Institute for Theoretical Physics, 31 Caroline 
Street North, Waterloo, ON, N2L 2Y5, Canada.\\
$^{2}$ Department of Physics and Astronomy, McMaster University, Hamilton, 
ON, L8S 4M1, Canada.}

\begin{abstract}

We argue that all necessary ingredients for successful inflation are present 
in the flat directions of the minimal supersymmetric standard model (MSSM). 
Out of many gauge invariant combinations of the squarks, sleptons and 
Higgses, there are two flat directions, ${\bf LLe}$, and ${\bf udd}$, which 
are promising candidates for the inflaton. The model predicts more than $10^3$ 
e-foldings with an inflationary scale of $H_{\rm inf} \sim {\cal O}(1-10)$ 
GeV, provides a tilted spectrum $0.92 \leq n_s \leq 1$ with an amplitude of 
$\delta_H \sim 10^{-5}$ and a negligible tensor perturbations. It yields a 
lower bound $\sim 340$ GeV on the sparticle masses which is within the reach 
of LHC. There is no gravitino or moduli problem in this model.        

\end{abstract}

\maketitle

\thispagestyle{fancy}


\section{Introduction}

The one crucial ingredient still missing in the otherwise highly successful 
theory of primordial inflation is its natural embedding within particle 
physics, particularly the standard model (SM) or its extesnions. In almost all 
models of inflation the inflaton is treated as s SM gauge singlet and, 
sometimes a complete gauge singlet whose origin nd couplings are chosen ad-hoc 
just to fit the observed cosmological data~\cite{WMAP3}.  

Recently we have constructed a model of inflation~\cite{AEGM,AEGJM} based on
the flat directions of MSSM (for a review of MSSM flat
directions, see~\cite{KARI-REV}). In this model the inflaton is a
gauge invariant combination of either squark or slepton fields.  For a
choice of the soft SUSY breaking parameters $A$ and the inflaton mass
$m_\phi$, the potential along the {\bf udd} and {\bf LLe}
directions is such that there is a period of slow roll inflation of
sufficient duration to provide the observed spectrum of CMB
perturbations
~\footnote{In a recent similar model with small Dirac
neutrino masses, the observed microwave background anisotropy and the
tilted power spectrum are related to the neutrino
masses~\cite{AKM}. The model relies solely on renormalizable
couplings.}.

MSSM inflation occurs at a very low scale with $H_{inf}\sim 1-10$~GeV
and with field values much below the Planck scale. Hence it stands in
strong contrast to the conventional inflation models which are based
on ad hoc gauge singlet fields and often employ field values close to
Planck scale (for a review, see ~\cite{LYTH}). In such models the
inflaton couplings to SM physics are unknown.  As a consequence, much
of the post-inflationary evolution, such as reheating, thermalization,
generation of baryon asymmetry and cold dark matter, which all depend
very crucially on how the inflaton couples to the (MS)SM
sector~\cite{AVERDI1,AVERDI2,AVERDI3,AVERDI4}, is not calculable from
first principles. The great virtue of MSSM inflation based on flat
directions is that the inflaton couplings to SM particles
are known. More importantly,  the inflaton mass is related either to squark or 
slepton masses, it could be measured by LHC or a future Linear Collider.

\section{The Model}\label{model}

Let us recapitulate the main features of MSSM flat direction
inflation~\cite{AEGM,AEGJM}. As is well known, in the limit of unbroken SUSY
the flat directions have exactly vanishing potential. This situation
changes if we take into account soft SUSY breaking and
non-renormalizable superpotential terms
of the type~\cite{KARI-REV}
\beq \label{supot}
W_{non} = \sum_{n>3}{\lambda_n \over n}{\Phi^n \over M^{n-3}}\,,
\eeq
where $\Phi$ is a
superfield which contains the
flat direction.  Within MSSM all the flat directions are lifted by
non-renormalizable operators with $4\le n\le 9$~\cite{GKM}. We expect that 
quantum gravity effects yield
$M=M_{\rm P}=2.4\times
10^{18}$~GeV and $\lambda_n\sim {\cal
O}(1)$~\cite{DRT}
.




Let us focus on the lowest order superpotential term in
Eq.~(\ref{supot}) which lifts the flat direction. Soft SUSY breaking
induces a mass term for $\phi$ and an $A$-term so that the scalar
potential along the flat direction reads
\beq \label{scpot}
V = {1\over2} m^2_\phi\,\phi^2 + A\cos(n \theta  + \theta_A)
{\lambda_{n}\phi^n \over n\,M^{n-3}_{\rm P}} + \lambda^2_n
{{\phi}^{2(n-1)} \over M^{2(n-3)}_{\rm P}}\,,
\eeq
Here $\phi$ and $\theta$ denote respectively the radial and the
angular coordinates of the complex scalar field
$\Phi=\phi\,\exp[i\theta]$, while $\theta_A$ is the phase of the
$A$-term (thus $A$ is a positive quantity with dimension of mass).
Note that the first and third terms in Eq.~(\ref{scpot}) are positive
definite, while the $A$-term leads to a negative contribution along
the directions whenever $\cos(n \theta + \theta_A) < 0$.~\footnote{The
importance of the A-term has also been highlighted in a successful
MSSM curvaton model~\cite{AEJM}.}




The maximum impact from the $A$-term is obtained when $\cos(n \theta +
\theta_A) = -1$ (which occurs for $n$ values of $\theta$).  
%
%


In the gravity mediated SUSY breaking case, the $A$-term and
the soft SUSY breaking mass terms are expected to be the same order of
magnitude as the gravitino mass, i.e. $m_{\phi}\sim A \sim m_{3/2}\sim 
{\cal O}(1)~{\rm TeV}$.
Now if $A$ and $m_\phi$ are related by
\beq
\label{cond}
A^2 = 8 (n-1) m^2_\phi\,,
\eeq
%
then both the first and second
derivatives of $V$ vanish at $\phi_0$, ~i.e. $V^{\prime}(\phi_0)=0,~
V^{\prime\prime}(\phi_0)=0$, where
\beq \label{phi0}
\phi_0 = \left({m_\phi M^{n-3}_{\rm P}\over
\lambda_n\sqrt{2n-2}}\right)^{1/(n-2)}\,.
\eeq
%




The potential near the saddle point Eq.~(\ref{phi0}) is very flat
along the {\it real direction} but not along the {\it imaginary
direction}. Along the imaginary direction the curvature is determined
by $m_{\phi}$.  Around $\phi_0$ the field lies in a plateau with a
potential energy
\beq \label{potential}
V(\phi_0) = {(n-2)^2\over2n(n-1)}\,m^2_\phi \phi_0^2.
\eeq
As the result, if initially $\phi \sim
\phi_0$, a slow roll phase of inflation is driven by the third
derivative of the potential.
The Hubble expansion rate during inflation which is given by
\beq \label{hubble}
H_{\rm inf} = {(n-2) \over \sqrt{6 n (n-1)}} {m_{\phi} \phi_0 \over M_{\rm P}}.
\eeq
When $\phi$ is very close to $\phi_0$, the first derivative is
extremely small. The field is effectively in a de Sitter background,
and we are in self-reproduction (or eternal inflation) regime where the
two point correlation function for the flat direction fluctuation
grows with time. But eventually classical friction wins and slow roll
begins at $\phi \approx \phi_{\rm self}$~\cite{AEGM}
\beq \label{self}
(\phi_0-\phi_{\rm self}) \simeq \Big({m_\phi \phi_0^2 \over M_{\rm
P}^3}\Big)^{1/2} \phi_0.
\eeq
The slow roll potential in this case reads
\begin{eqnarray} \label{potential2}
&&V(\phi) = V(\phi_0) + {1\over3!} V'''(\phi_0)
(\phi-\phi_0)^3 +\cdot\cdot\cdot \,, \nonumber \\
\label{3rder}
&&V^{\prime \prime \prime}({\phi_0}) = 2(n-2)^2
{m^2_\phi \over \phi_0}\,.
\end{eqnarray}
We can now solve the equation of motion for the $\phi$ field in the
slow-roll approximation,
\beq \label{slow}
3H\dot\phi=-\frac{1}{2}V'''(\phi_0)(\phi-\phi_0)^2\,,
\eeq
assuming initial conditions such that the flat direction starts in the
vicinity of $\phi_0$ with $\dot\phi\approx 0$.  Inflation ends when
either of the slow roll parameters, $\epsilon\equiv (M_{\rm
P}^2/2)(V^{\prime}/V)^2$ and $\eta \equiv M^2_{\rm P}(V^{\prime
\prime}/V)$, becomes of ${\cal O}(1)$. It happens that $\vert \eta
\vert \sim 1$ when $\phi \approx \phi_{\rm end}$, where
\beq \label{end}
(\phi_0-\phi_{\rm end}) \sim {\phi^3_0 \over 4n(n-1)M^2_{\rm P}}\,.
\eeq
The number of e-foldings during the slow roll from $\phi$ to
$\phi_{\rm end}$ is given by
\beq \label{efold}
{\cal N}_e(\phi) = \int_{\phi}^{\phi_{\rm end}} {H_{\rm inf} d\phi
\over \dot\phi} \simeq {\phi^3_0 \over 2n(n-1)M^2_{\rm P}(\phi_0 - \phi)}\,,
\eeq
where we have used $V'(\phi) \sim (\phi - \phi_0)^2 V'''(\phi_0)$
(this is justified since $V'(\phi_0) \sim 0, V''(\phi_0)\sim 0$), and
Eq.~(\ref{slow}). The total number of e-foldings in the slow roll
regime is then found from Eq.(\ref{self})
\beq \label{tot}
{\cal N}_{\rm tot} \simeq {1 \over 2n(n-1)} \big({\phi^2_0 \over m_{\phi}
M_{\rm P}}\Big)^{1/2}.
\eeq
The observationally relevant perturbations are generated when $\phi
\approx \phi_{\rm COBE}$. The number of e-foldings between $\phi_{\rm
COBE}$ and $\phi_{\rm end}$, denoted by ${\cal N}_{\rm COBE}$ follows
from Eq.~(\ref{efold})
\beq \label{cobe}
{\cal N}_{\rm COBE} \simeq {\phi^3_0 \over 2n(n-1)M^2_{\rm P}(\phi_0 -
\phi_{\rm COBE})}.
\eeq
The amplitude of perturbations thus produced is given by
\beq \label{ampl}
\delta_{H} \equiv \frac{1}{5\pi}\frac{H^2_{\rm inf}}{\dot\phi} \simeq
\frac{1}{5\pi} \sqrt{\frac{2}{3}n(n-1)}(n-2) ~ \Big({m_\phi M_{\rm P} \over
\phi_0^2}\Big) ~ {\cal N}_{\rm COBE}^2,
\eeq
where we have used
Eqs.(\ref{hubble},\ref{potential2},\ref{cobe}). Again after using
these equations, the spectral tilt of the power spectrum and its
running are found to be
\begin{eqnarray}
\label{tilt}
&&n_s = 1 + 2\eta - 6\epsilon \simeq 1 -
{4\over {\cal N}_{\rm COBE}}, \, \\ \nonumber
&&{d\,n_s\over d\ln k} = - {4\over {\cal N}_{\rm COBE}^2}. \,
\end{eqnarray}
%

\section{Properties and predictions}

As discussed in~\cite{AEGM}, among the about 300 flat directions there
are two that can lead to a successful inflation along the lines
discussed above.

One is {\bf udd} which, up to an overall phase factor, is parameterized by
\beq
\label{example}
u^{\alpha}_i=\frac1{\sqrt{3}}\phi\,,~
d^{\beta}_j=\frac1{\sqrt{3}}\phi\,,~
d^{\gamma}_k=\frac{1}{\sqrt{3}}\phi\,.
\eeq
Here $1 \leq \alpha,\beta,\gamma \leq 3$ are color indices, and $1
\leq i,j,k \leq 3$ denote the quark families. The flatness constraints
require that $\alpha \neq \beta \neq \gamma$ and $j \neq k$.

The other direction is {\bf LLe}, parameterized by (again up to an overall
phase factor)
\beq
L^a_i=\frac1{\sqrt{3}}\left(\begin{array}{l}0\\ \phi\end{array}\right)\,,~
L^b_j=\frac1{\sqrt{3}}\left(\begin{array}{l}\phi\\ 0\end{array}\right)\,,~
e_k=\frac{1}{\sqrt{3}}\phi\,,
\eeq
where $1 \leq a,b \leq 2$ are the weak isospin indices and $1 \leq
i,j,k \leq 3$ denote the lepton families. The flatness constraints
require that $a \neq b$ and $i \neq j \neq k$.  Both these flat
directions are lifted by $n=6$ non-renormalizable operators,
\begin{eqnarray}
W_6\supset\frac{1}{M_{\rm P}^3}(LLe)(LLe)\,,\hspace{1cm}
W_6\supset \frac{1}{M_{\rm P}^3}(udd)(udd)\,.
\end{eqnarray}
The reason for choosing either of these two flat
directions
is twofold:
(i) a non-trivial $A$-term arises, at the lowest order, only at $n=6$;
and (ii) we wish to obtain the correct COBE normalization of the CMB
spectrum 9for details, see~\cite{AEGJM}).

%
%


According to the arguments presented above, successful MSSM flat direction
inflation has the following model parameters:
\beq
m_{\phi}\sim 1-10~{\rm TeV}\,,~~n=6\,,~~A=\sqrt{40}m_{\phi}\,,
~~\lambda\sim {\cal O}(1)\,.
\label{VALVS}
\eeq
Here we assume that $\lambda$ (we drop the subscript "6") is of order
one, which is the most natural assumption when $M=M_P$.

The Hubble expansion rate during inflation and the VEV of the saddle
point are
%
\beq \label{values}
H_{\rm inf}\sim 1-10~{\rm GeV}\,,~~~\phi_0 \sim (1-3) \times
10^{14}~{\rm GeV}\,.
\eeq
Note that both the scales are sub-Planckian. The total energy density
stored in the inflaton potential is $V_0 \sim 10^{36}-10^{38}~{\rm
GeV}^4$. The fact that $\phi_0$ is sub-Planckian guarantees that the
inflationary potential is free from the uncertainties about physics at
super-Planckian VEVs. The total number of e-foldings during the slow
roll evolution is large enough to dilute any dangerous relic away, see
Eq.~(\ref{tot}):
\beq \label{totalefold}
{\cal N}_{\rm tot} \sim 10^3  \,,
\eeq
Domains which are initially closer than $\phi_{\rm self}$ to $\phi_0$,
see Eq.~(\ref{self}), can enter self-reproduction in eternal
inflation, with no observable consequences.

At such low scales as in MSSM inflation the number of e-foldings,
${\cal N}_{\rm COBE}$, required for the observationally relevant
perturbations, is much less than $60$~\cite{MULTI}.  If the inflaton
decays immediately after the end of inflation, we obtain ${\cal
N}_{\rm COBE} \sim 50$. Despite the low scale, the flat direction can
generate adequate density perturbations as required to explain the
COBE normalization. This is due to the extreme flatness of the
potential (recall that $V'=0$), which causes the velocity of the
rolling flat direction to be extremely small. From Eq.~(\ref{ampl}) we
find an amplitude of
\beq
\label{amp}
\delta_{H} \simeq 1.91 \times 10^{-5}\,.
\eeq

There is a constraint on the mass of the flat direction from the
amplitude of the CMB anisotropy:
\begin{equation}
\label{mbound}
m_{\phi} = (340 ~ {\rm GeV}) \times \lambda^{-1} \,
\left( \frac{{\cal N}_{\rm COBE}}{50} \right)^{-4}\,.
\end{equation}
We get a lower limit on the mass parameter when $\lambda\leq 1$.
For smaller values of $\lambda\ll 1$, the mass of the flat
direction must be larger.  Note that the above bound on the inflaton
mass arises at high scales, i.e. $\phi=\phi_0$. However, through
renormalization group flow, it is connected to the low scale mass~\cite{AEGJM}.
               
The spectral tilt of the power spectrum is not negligible because,
although the first slow roll parameter is $\epsilon\sim1/{\cal N}_{\rm
COBE}^4\ll 1$, the other slow roll parameter is given by $\eta =
-2/{\cal N}_{\rm COBE}$ and thus, see
Eq.~(\ref{tilt})\footnote{Obtaining $n_s > 0.92$ requires deviation
from the saddle point condition in Eq.~(\ref{cond}). Recently we have
illustrated that within MSSM inflation the spectral tilt varies from
$0.92 \leq n_s\leq 1$ ~\cite{Maz-Roz}.  Inflation occurring exactly at
the saddle point yields the lower limit, while the upper limit is
obtained when the total number of e-foldings is saturated by the
number of e-foldings required for the COBE normalization, i.e. ${\cal
N}_{\rm tot}\approx {\cal N}_{\rm COBE}\sim 50$. }
\begin{eqnarray}
\label{spect}
&&n_s
\sim 0.92\,,\\
&&{d\,n_s\over d\ln k}
\sim - 0.002\,,
\end{eqnarray}
where we have taken ${\cal N}_{\rm COBE} \sim 50$ (which is the
maximum value allowed for the scale of inflation in our
model)
. In the absence of tensor
modes, this agrees with the current WMAP 3-years' data within
$2\sigma$~\cite{WMAP3}. Note that MSSM inflation does not produce any
large stochastic gravitational wave background during
inflation. Gravity waves depend on the Hubble expansion rate, and in
our case the energy density stored in MSSM inflation is very small.

\section{Sensitivity of MSSM inflation}

The dynamics of the flat
direction inflaton has been discussed assuming the saddle point condition
Eq.~(\ref{cond}) is satisfied exactly.  The question then is, how
large a deviation can be allowed for before slow roll inflation will
be spoiled. To facilitate the discussion, let us define
\begin{eqnarray}
\delta\equiv {A^2\over40m_\phi^2}
\equiv 1 \pm 4\,\alpha^2\,,
\end{eqnarray}
where $\alpha \ll 1$. There are two distinct possibilities: either
$\delta > 1$ or $\delta < 1$ (when $\delta = 1$ we recover the saddle point 
condition). In the former case there is a 
barrier which separates the global minimum $\phi=0$ and the false minimum at
$\phi\simeq\phi_0$. If the barrier is too high, the field remains
trapped and there is no slow-roll inflation. Therefore one must
require that tunneling is effective
so that the field can jump to the top of the barrier.  In the latter
case there is no minimum but the potential may be too steep for slow
roll inflation. 

Obtaining sufficient inflation, i.e. such that the number of e-foldings is 
$\geq {\cal N}_{\rm COBE}$, with observationally acceptable properties
results in a
following constraint on $\alpha$,
\beq \label{slowroll2}
\alpha \leq {1 \over 30 {\cal N}_{\rm COBE}}
\Big({\phi_0 \over M_{\rm P}}\Big)^2,
\eeq
in both cases with $\delta > 1$ and $\delta < 1$~\cite{AEGJM}. For typical 
values of $\phi_0 \lsim
10^{15}$~GeV
, the saddle point condition in Eq.~(\ref{cond}) requires tuning at a level;
\beq \label{fine}
\alpha \sim 10^{-9}\,.
\eeq

Since radiative corrections modify $\alpha$, we need to finetune the potential 
to a few (but not all) orders in perturbation theory. Althoug not disastruous, 
this can hardly be considered a satisfactory situation. However, it is 
conceivable that the mechnism of supersymmetry breaking could remove the 
fine-tuning in some natural, dynamical way. For instance, $A/m$ could turn out 
to be a renormalization group (RG) fixed point so that once the ratio is 
fixed, it 
would remain fixed under quantum corrections (for example, 
see~\cite{ROSS}). This is an interesting possibility which requires a detailed 
investigation.

\section{The inflaton and LHC}

Let us recall that the constraint on the mass of the $n=6$ flat
direction inflaton in Eq.~(\ref{mbound}).
As mentioned earlier, this is the bound on the mass of the flat
direction during inflation, determined at the scale $\phi=\phi_0$.
Since the inflaton mass runs from $\phi_0$ down to the LHC energy
scales, it will also get scaled.

For ${\bf LLe}$ the flat direction mass only gets larger due to the gaugino 
loops. As a result, the lower bound on the slepton masses is larger by 
$\leq 50\%$ compared to that in Eq.~(\ref{mbound})~\cite{AEGJM}). 
The situation would be similar for ${\bf udd}$ without the top squark.
For the ${\bf u_3dd}$ direction it is possible that the inflaton mass
gets even smaller at the weak scale.
We do not claim
that LHC can discover MSSM inflation, but it can certainly rule out
the possibility. If LHC does not find low energy supersymmetry within
$\sim $~TeV, then MSSM inflation is effectively ruled out.

Unlike
$m_{\phi}$, there is no prospect of measuring the $A$ term, because it
is related to the non-renormalizable interactions which are suppressed
by $M_{\rm P}$. However, a knowledge of supersymmetry breaking sector
and its communication with the observable sector may help to link the
non-renormalizable $A$-term under consideration to the renormalizable
ones.

To elucidate this, let us consider the Polonyi model where a general
$A$-term at a tree level is given by
$$m_{3/2}[(a-3)W+\phi (dW/d\phi)],$$
with $a=3 - \sqrt{3}$~\cite{NILLES}. One then finds a relationship
between $A$-terms corresponding to $n=6$ and $n=3$ superpotential
terms, denoted by $A_6$ and $A_3$ respectively, at high scales:
\beq \label{polon}
A_6={3 - \sqrt{3} \over 6 - \sqrt{3}} A_3\,.
\eeq
One can then use relevant RG equations to relate $A_6$ which is relevant for
inflation, to $A_3$ at the weak scale, which can be constrained and/or
measured. In principle this can also be done in general, provided that
we have sufficient information about the supersymmetry breaking sector
and its communication with the MSSM sector.

\section{End of MSSM inflation}

After the end of inflation, the flat direction starts rolling towards
its global minimum. At this stage the dominant term in the scalar
potential will be: $m_\phi \phi^2/2$. Since the frequency of
oscillations is $\omega \sim m_{\phi} \sim 10^3 H_{\rm inf}$, the flat
direction oscillates a large number of times within the first Hubble
time after the end of inflation. Hence the effect of expansion is
negligible.

We recall that the curvature of the potential along the angular
direction is much larger than $H^2_{\rm inf}$. Therefore, the flat
direction has settled at one of the minima along the angular direction
during inflation from which it cannot be displaced by quantum
fluctuations. This implies that no torque will be exerted, and hence
the flat direction motion will be one dimensional, i.e. along the
radial direction.

Flat direction oscillations excite those MSSM degrees of freedom which
are coupled to it.  The inflaton, either ${\bf LLe}$ or ${\bf u}{\bf
d}{\bf d}$ flat direction, is a linear combination of slepton or
squark fields. Therefore it has gauge couplings to the
gauge/gaugino fields 
and Yukawa couplings to the Higgs/Higgsino
fields. 
Let us elucidate the physics, by
considering the case when ${\bf LLe}$ flat direction is the inflaton
.




An efficient bout of particle creation
occurs when the inflaton crosses the origin, which happens twice in
every oscillation. The reason is that fields which are coupled to the
inflaton are massless near the point of enhanced symmetry. Mainly
electroweak gauge fields and gauginos are then created as they have
the largest coupling to the flat direction. 
The production takes place
in a short interval, $\Delta t \sim \left(g m_{\phi} \phi_0
\right)^{-1/2}$, where $\phi_0\sim 10^{14}$~GeV is the initial
amplitude of the inflaton oscillation, during which quanta with a
physical momentum $k \lsim \left(g m_{\phi} \phi_0 \right)^{1/2}$ are
produced. 
%
%
As the inflaton VEV is rolling back to its maximum value $\phi_0$, the
mass of the produced quanta $g \langle \phi(t) \rangle$ increases. The
gauge and gaugino fields can (perturbatively) decay to the fields
which are not coupled to the inflaton, for instance to (s)quarks. Note
that (s)quarks are not coupled to the flat direction, hence they
remain massless throughout the oscillations. 

%
%
The decay is very quick
compared with the frequency of inflaton oscillations
. 
%
%
The ratio of energy density in relativistic particles thus produced
$\rho_{rel}$ with respect to the total energy density $\rho_0$ is
%
\beq
\label{ratio}
{\rho_{rel} \over \rho_0} \sim 10^{-2} g\,.
\eeq
%
This implies that a
fraction $\sim {\cal O}(10^{-2})$ of the inflaton energy density is
transferred into relativistic (s)quarks every time that the inflaton
passes through the origin. This is so-called instant preheating
mechanism~\cite{INSTANT}.
It is quite an efficient mechanism in our model
as it can convert almost all of the energy density in the inflaton into
radiation within a Hubble time 



The maximum temperature attained by the plasma would be given by:
\beq
\label{tmax}
T_{max} \sim \left(m_{\phi}\phi_0\right)^{1/2}
\geq 10^{9}~{\rm GeV}\,.
\eeq
This temperature may be too high and could lead to thermal
overproduction of gravitinos~\cite{Ellis,Buchmuller}. However the
dominant source of gravitino production in a thermal bath is
scatterings which include an on-shell gluon or gluino leg. 



In order to suppress thermal gravitino production it is therefore 
sufficient to make gluon and gluino fields heavy enough such that they are
not kinematically accessible to the reheated plasma. 
This suggests a natural solution to the thermal gravitino problem
in the
case of our model. Consider another flat direction with a
non-zero VEV, denoted by $\varphi$, which spontaneously breaks the
$SU(3)_C$ group.  For example, if ${\bf LLe}$ is the inflaton, then
${\bf udd}$ provides a unique candidate which can simultaneously
develop VEV.
%
%

So long as $g \varphi \gg T$, the gluon/gluino fields will be too 
heavy
and not kinematically accessible to the reheated plasma. 
Once $g \langle \varphi \rangle \simeq T$,
gluon/gluino fields come into equilibrium with the thermal bath.  As
pointed out in Refs.~\cite{AVERDI1,AVERDI2}, if the initial VEV of
${\bf udd}$ is
%
$\varphi_0 > 10^{10}~{\rm GeV}$,
%
then the temperature at which gluon/gluino become kinematically
accessible, i.e. $g \langle \varphi \rangle \simeq T$, is given by
~\cite{AVERDI2}
%
\beq
T_{\rm R} \leq 10^{7}~{\rm GeV}\,.
\eeq
This is the final reheat temperature at which gluons and
gluinos are all in thermal equilibrium with the other degrees
of freedom. The standard calculation of thermal gravitino production via
scatterings can then be used for $T \leq T_{\rm R}$. Note however that 
$T_{\rm R}$
is sufficiently low to avoid thermal overproduction of gravitinos.

Finally, we also make a comment on the cosmological moduli problem.
The moduli are generically displaced from their true
minimum if their mass is less than the expansion rate during
inflation. The moduli obtain a mass $\sim {\cal O}({\rm TeV})$ from
supersymmetry breaking. They start oscillating with a large amplitude,
possibly as big as $M_{\rm P}$, when the Hubble parameter drops below
their mass. Since moduli are only gravitationally coupled to other
fields, their oscillations dominate the Universe while they decay very
late.  The resulting reheat temperature is below MeV, and is too low
to yield a successful primordial nucleosynthesis.

However, in our case $H_{\rm inf} \ll {\rm TeV}$ . This
implies that quantum fluctuations cannot displace the moduli from
their true minima during the inflationary epoch driven by MSSM flat
directions. Moreover, any oscillations of the moduli will be
exponentially damped during the inflationary epoch. Therefore our
model is free from the infamous moduli problem.

\section{Conclusion}

The existence of a saddle point in the scalar potential of the ${\bf
udd}$ or ${\bf LLe}$ MSSM flat directions appears, perhaps
surprisingly, to provide all the necessary ingredients for an
observationally realistic model of inflation. 
The exceptional
feature of the model, which sets it apart from conventional singlet
field inflation models, is the fact that here the inflaton is a gauge
invariant combination of the squark or slepton fields. 
As a consequence, the mass of the inflaton is not a free
parameter but is related to the masses of e.g. sleptons, should the
${\bf LLe}$ direction be the inflaton. Hence LHC can indeed put a constraint 
on the model: it may not be able
to verify it, but it certainly can rule it out.
Moreover, the couplings of the inflaton to the MSSM matter and
gauge fields are known.  This makes it possible to address the
questions of reheating in an unambiguous way.
As we saw, the model is free from the gravitino and moduli problems.

To summarize, MSSM flat
direction inflation is unique in being both a successful
model of inflation and at the same time having a concrete and real
connection to physics that can be observed in earth bound
laboratories.

\bigskip
\begin{acknowledgments}

I am indebted to my collaborators Kari Enqvist, Juan Garcia-Bellido, 
Asko Jokinen, Alexander Kusenko and Anupam Mazumdar. I also thank the 
organizers of IPM-LHP06 for their invitation and kind hospitality. This work
was supported by the National Sciences and Engineering Research Council of 
Canada (NSERC) and by Perimeter Institute for Theoretical Physics. Research
at Perimetre Institute is supported in part by the province of Ontario through 
MEDT.

\end{acknowledgments}

\bigskip 

\end{document}